\documentclass[RNAAS]{aastex63}
\usepackage{amsmath,amssymb,CJK}

\begin{document}
\begin{CJK*}{UTF8}{gbsn}

\title{Recurring Outbursts of P/2019 LM$_4$ (Palomar)}

\correspondingauthor{Quanzhi Ye}
\email{qye@umd.edu}

\author[0000-0002-4838-7676]{Quanzhi Ye (叶泉志)}
\affiliation{Department of Astronomy, University of Maryland, College Park, MD 20742, USA}

\author[0000-0002-6702-7676]{Michael S. P. Kelley}
\affiliation{Department of Astronomy, University of Maryland, College Park, MD 20742, USA}

\author[0000-0002-2668-7248]{Dennis Bodewits}
\affiliation{Physics Department, Auburn University, Auburn, AL 36849, USA}

\author[0000-0002-6702-7676]{James M. Bauer}
\affiliation{Department of Astronomy, University of Maryland, College Park, MD 20742, USA}

\author[0000-0003-2242-0244]{Ashish Mahabal}
\affiliation{Division of Physics, Mathematics, and Astronomy, California Institute of Technology, Pasadena, CA 91125, USA}
\affiliation{Center for Data Driven Discovery, California Institute of Technology, Pasadena, CA 91125, USA}

\author[0000-0002-8532-9395]{Frank J. Masci}
\affiliation{IPAC, California Institute of Technology, 1200 E. California Blvd, Pasadena, CA 91125, USA}

\author[0000-0001-8771-7554]{Chow-Choong Ngeow}
\affiliation{Institute of Astronomy, National Central University, 32001, Taiwan, R.O. China}


\begin{abstract}
    We present a preliminary analysis of comet P/2019 LM$_4$ (Palomar) as observed by the Zwicky Transient Facility (ZTF) survey in 2019 and 2020. We find that the discovery of the comet in 2019 and the recovery in 2020 is largely attributed to two separate outbursts that are $\gtrsim2$ and $\gtrsim3.9$ mag in strength. The outbursts occurred around the end of April to early May of 2019 as well as between 2020 May 8.31 and 9.52 UTC, respectively.
\end{abstract}


\section{}

Cometary outbursts are brief, unexpected increases in the brightness of comets that indicate a sudden and dramatic change in the activity \citep{Patashnick1974, Hughes1990}. Such events sometimes lead to the discovery of originally inactive or weakly active comets which would otherwise not be detectable \citep{Kramer2017, Ye2017}.

Our preliminary analysis of images of comet P/2019 LM$_4$ (Palomar) acquired with the Zwicky Transient Facility (ZTF) survey \citep{Bellm2019} has revealed recurring brightenings or outbursts of activity. P/Palomar was first detected as an asteroidal object on 2019 June 4 and 7 in the course of the ZTF Twilight Survey \citep{Ye2020}. Following the recent recovery of the object\footnote{See Minor Planet Electronic Circular 2020-J68: \url{https://www.minorplanetcenter.net/mpec/K20/K20J68.html}, as well as Central Bureau Electronic Telegram 4775: \url{http://www.cbat.eps.harvard.edu/iau/cbet/004700/CBET004775.txt}.}, we searched the ZTF Data Release 2 \citep{Masci2019} and Partnership archives within the $3\sigma$ positional uncertainty as predicted by JPL orbit solution \#3 for additional observations. The data are calibrated to the Pan-STARRS 1 photometric system \citep{Masci2019}, with colors corrected assuming a reddened solar spectrum ($g_\mathrm{P1}-r_\mathrm{P1}=0.49$ and $r_\mathrm{P1}-i_\mathrm{P1}=0.34$). For images where the comet is detected, the photometry is measured using a $5''$ radius aperture. For the non-detections, we compute the $5\sigma$ upper limits based on the ZTF pipeline.

\begin{figure}[h!]
\begin{center}
\includegraphics[scale=1,angle=0]{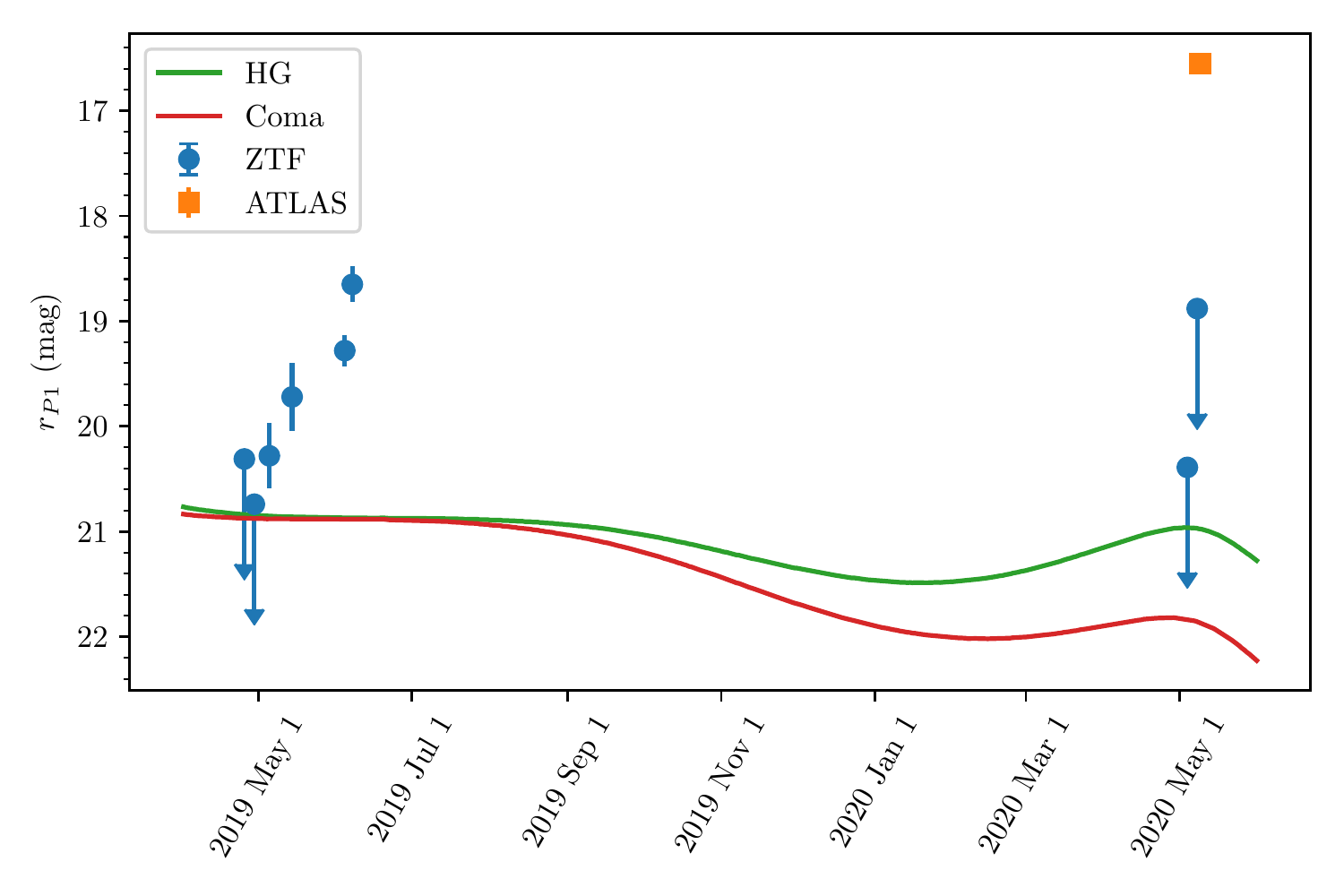}
\caption{The composite lightcurve of P/2019 LM$_4$ made using the ZTF detections and $5\sigma$ upper limits. The ATLAS detection on 2020 May 9 at $r_\mathrm{P1}=16.8$ is also shown for reference. Also shown are the HG model and a coma model computed using the Halley--Marcus phase function. The ZTF photometry is available as the Data behind the Figure. \label{fig:lc}}
\end{center}
\end{figure}

Figure~\ref{fig:lc} shows the composite lightcurve of P/Palomar produced using ZTF observations. The ATLAS precovery on 2020 May 9 is also plotted for reference, with an assumed color of $r_\mathrm{P1}-o_\mathrm{ATLAS}=0.05$ derived from a solar spectrum and the filter throughput curves \citep{astm06, Tonry2012, Tonry2018}. We also plot two lightcurve models: the HG model \citep{Bowell1989} assumes $H=14.1$ and $G=0.15$, while the coma model is computed using the Halley--Marcus phase function \citep{Schleicher1998, Marcus2007, Schleicher2011} following $m_\mathrm{coma} = 14.1 + 5 \log{\varDelta} + 10 \log{r_\mathrm{H}} - 2.5 \log{\Phi{(\alpha)}}$, where $\varDelta$ and $r_\mathrm{H}$ is geocentric and heliocentric distances in au, and $\Phi{(\alpha)}$ is the Halley--Marcus phase function applied to phase angle $\alpha$.

Precovery detections are identified on images taken on 2019 May 5 and 14, but not on images taken on 2019 April 29 or earlier. It can be concluded that the comet had brightened by at least 2 mag around the end of April to early May of 2019, when it was at $\sim2.4$ au from the Sun or $\sim50$ days before perihelion. As shown in Figure~\ref{fig:lc}, a normal comet would have largely remained at a constant apparent magnitude over this period. In 2020, the most recent pre-discovery ZTF field that covers the uncertainty ellipse of the comet, taken on UT 2020 May 8.31, reveals no comet. The first detection was made by the Asteroid Terrestrial-Impact Last Alert System (ATLAS) program \citep{Tonry2018} on UT 2020 May 9.52 UTC. This suggests an outburst onset between May 8.31 and 9.52 UTC, with a strength of at least 3.9 mag, corresponding to a phase-corrected $Af\rho \approx 16 \pm 2$~cm. The outburst happened when the comet was at a heliocentric distance of 3.50 au, 325 to 326 days after perihelion.

\acknowledgments

Based on observations obtained with the Samuel Oschin Telescope 48-inch at the Palomar Observatory as part of the Zwicky Transient Facility project. ZTF is supported by the National Science Foundation under Grant No. AST-1440341 and a collaboration including Caltech, IPAC, the Weizmann Institute for Science, the Oskar Klein Center at Stockholm University, the University of Maryland, the University of Washington, Deutsches Elektronen-Synchrotron and Humboldt University, Los Alamos National Laboratories, the TANGO Consortium of Taiwan, the University of Wisconsin at Milwaukee, and Lawrence Berkeley National Laboratories. Operations are conducted by COO, IPAC, and UW.

\software{{\tt sbpy} \citep{Mommert2019}, {\tt ZChecker} \citep{Kelley2019}}
\facilities{PO:1.2m}

\bibliographystyle{aasjournal}
\bibliography{sample63}{}

\begin{thebibliography}{}
\expandafter\ifx\csname natexlab\endcsname\relax\def\natexlab#1{#1}\fi
\providecommand{\url}[1]{\href{#1}{#1}}
\providecommand{\dodoi}[1]{doi:~\href{http://doi.org/#1}{\nolinkurl{#1}}}
\providecommand{\doeprint}[1]{\href{http://ascl.net/#1}{\nolinkurl{http://ascl.net/#1}}}
\providecommand{\doarXiv}[1]{\href{https://arxiv.org/abs/#1}{\nolinkurl{https://arxiv.org/abs/#1}}}

\bibitem[{{ASTM International}(2006)}]{astm06}
{ASTM International}. 2006, ASTM Standard E490, 2000 (2006), Solar Constant and
  Zero Air Mass Solar Spectral Irradiance Tables (West Conshohocken, PA: ASTM
  International), \dodoi{10.1520/E0490-00AR06}

\bibitem[{{Bellm} {et~al.}(2019){Bellm}, {Kulkarni}, {Graham}, {Dekany},
  {Smith}, {Riddle}, {Masci}, {Helou}, {Prince}, {Adams}, {Barbarino},
  {Barlow}, {Bauer}, {Beck}, {Belicki}, {Biswas}, {Blagorodnova}, {Bodewits},
  {Bolin}, {Brinnel}, {Brooke}, {Bue}, {Bulla}, {Burruss}, {Cenko}, {Chang},
  {Connolly}, {Coughlin}, {Cromer}, {Cunningham}, {De}, {Delacroix}, {Desai},
  {Duev}, {Eadie}, {Farnham}, {Feeney}, {Feindt}, {Flynn}, {Franckowiak},
  {Frederick}, {Fremling}, {Gal-Yam}, {Gezari}, {Giomi}, {Goldstein},
  {Golkhou}, {Goobar}, {Groom}, {Hacopians}, {Hale}, {Henning}, {Ho}, {Hover},
  {Howell}, {Hung}, {Huppenkothen}, {Imel}, {Ip}, {Ivezi{\'c}}, {Jackson},
  {Jones}, {Juric}, {Kasliwal}, {Kaspi}, {Kaye}, {Kelley}, {Kowalski},
  {Kramer}, {Kupfer}, {Landry}, {Laher}, {Lee}, {Lin}, {Lin}, {Lunnan},
  {Giomi}, {Mahabal}, {Mao}, {Miller}, {Monkewitz}, {Murphy}, {Ngeow},
  {Nordin}, {Nugent}, {Ofek}, {Patterson}, {Penprase}, {Porter}, {Rauch},
  {Rebbapragada}, {Reiley}, {Rigault}, {Rodriguez}, {van Roestel}, {Rusholme},
  {van Santen}, {Schulze}, {Shupe}, {Singer}, {Soumagnac}, {Stein}, {Surace},
  {Sollerman}, {Szkody}, {Taddia}, {Terek}, {Van Sistine}, {van Velzen},
  {Vestrand}, {Walters}, {Ward}, {Ye}, {Yu}, {Yan}, \& {Zolkower}}]{Bellm2019}
{Bellm}, E.~C., {Kulkarni}, S.~R., {Graham}, M.~J., {et~al.} 2019, \pasp, 131,
  018002, \dodoi{10.1088/1538-3873/aaecbe}

\bibitem[{{Bowell} {et~al.}(1989){Bowell}, {Hapke}, {Domingue}, {Lumme},
  {Peltoniemi}, \& {Harris}}]{Bowell1989}
{Bowell}, E., {Hapke}, B., {Domingue}, D., {et~al.} 1989, in Asteroids II, ed.
  R.~P. {Binzel}, T.~{Gehrels}, \& M.~S. {Matthews}, 524--556

\bibitem[{{Hughes}(1990)}]{Hughes1990}
{Hughes}, D.~W. 1990, \qjras, 31, 69

\bibitem[{{Kelley} {et~al.}(2019){Kelley}, {Bodewits}, {Ye}, {Laher}, {Masci},
  {Monkewitz}, {Riddle}, {Rusholme}, {Shupe}, \& {Soumagnac}}]{Kelley2019}
{Kelley}, M. S.~P., {Bodewits}, D., {Ye}, Q., {et~al.} 2019, in Astronomical
  Society of the Pacific Conference Series, Vol. 523, Astronomical Data
  Analysis Software and Systems XXVII, ed. P.~J. {Teuben}, M.~W. {Pound}, B.~A.
  {Thomas}, \& E.~M. {Warner}, 471

\bibitem[{{Kramer} {et~al.}(2017){Kramer}, {Bauer}, {Fernandez}, {Stevenson},
  {Mainzer}, {Grav}, {Masiero}, {Nugent}, \& {Sonnett}}]{Kramer2017}
{Kramer}, E.~A., {Bauer}, J.~M., {Fernandez}, Y.~R., {et~al.} 2017, \apj, 838,
  58, \dodoi{10.3847/1538-4357/aa5f59}

\bibitem[{{Marcus}(2007)}]{Marcus2007}
{Marcus}, J.~N. 2007, International Comet Quarterly, 29, 39

\bibitem[{{Masci} {et~al.}(2019){Masci}, {Laher}, {Rusholme}, {Shupe}, {Groom},
  {Surace}, {Jackson}, {Monkewitz}, {Beck}, {Flynn}, {Terek}, {Landry},
  {Hacopians}, {Desai}, {Howell}, {Brooke}, {Imel}, {Wachter}, {Ye}, {Lin},
  {Cenko}, {Cunningham}, {Rebbapragada}, {Bue}, {Miller}, {Mahabal}, {Bellm},
  {Patterson}, {Juri{\'c}}, {Golkhou}, {Ofek}, {Walters}, {Graham}, {Kasliwal},
  {Dekany}, {Kupfer}, {Burdge}, {Cannella}, {Barlow}, {Van Sistine}, {Giomi},
  {Fremling}, {Blagorodnova}, {Levitan}, {Riddle}, {Smith}, {Helou}, {Prince},
  \& {Kulkarni}}]{Masci2019}
{Masci}, F.~J., {Laher}, R.~R., {Rusholme}, B., {et~al.} 2019, \pasp, 131,
  018003, \dodoi{10.1088/1538-3873/aae8ac}

\bibitem[{{Mommert} {et~al.}(2019){Mommert}, {Kelley}, {de Val-Borro}, {Li},
  {Guzman}, {Sip{\H{o}}cz}, {{\v{D}}urech}, {Granvik}, {Grundy}, {Moskovitz},
  {Penttil{\"a}}, \& {Samarasinha}}]{Mommert2019}
{Mommert}, M., {Kelley}, M., {de Val-Borro}, M., {et~al.} 2019, The Journal of
  Open Source Software, 4, 1426, \dodoi{10.21105/joss.01426}

\bibitem[{{Patashnick}(1974)}]{Patashnick1974}
{Patashnick}, H. 1974, \nat, 250, 313, \dodoi{10.1038/250313a0}

\bibitem[{{Schleicher} \& {Bair}(2011)}]{Schleicher2011}
{Schleicher}, D.~G., \& {Bair}, A.~N. 2011, 141, 177,
  \dodoi{10.1088/0004-6256/141/6/177}

\bibitem[{{Schleicher} {et~al.}(1998){Schleicher}, {Millis}, \&
  {Birch}}]{Schleicher1998}
{Schleicher}, D.~G., {Millis}, R.~L., \& {Birch}, P.~V. 1998, 132, 397,
  \dodoi{10.1006/icar.1997.5902}

\bibitem[{Tonry {et~al.}(2012)Tonry, Stubbs, Lykke, Doherty, Shivvers, Burgett,
  Chambers, Hodapp, Kaiser, Kudritzki, Magnier, Morgan, Price, \&
  Wainscoat}]{Tonry2012}
Tonry, J.~L., Stubbs, C.~W., Lykke, K.~R., {et~al.} 2012, 750, 99,
  \dodoi{10.1088/0004-637X/750/2/99}

\bibitem[{{Tonry} {et~al.}(2018){Tonry}, {Denneau}, {Heinze}, {Stalder},
  {Smith}, {Smartt}, {Stubbs}, {Weiland }, \& {Rest}}]{Tonry2018}
{Tonry}, J.~L., {Denneau}, L., {Heinze}, A.~N., {et~al.} 2018, \pasp, 130,
  064505, \dodoi{10.1088/1538-3873/aabadf}

\bibitem[{{Ye} {et~al.}(2020){Ye}, {Masci}, {Ip}, {Prince}, {Helou},
  {Farnocchia}, {Bellm}, {Dekany}, {Graham}, {Kulkarni}, {Kupfer}, {Mahabal},
  {Ngeow}, {Reiley}, \& {Soumagnac}}]{Ye2020}
{Ye}, Q., {Masci}, F.~J., {Ip}, W.-H., {et~al.} 2020, \aj, 159, 70,
  \dodoi{10.3847/1538-3881/ab629c}

\bibitem[{{Ye}(2017)}]{Ye2017}
{Ye}, Q.-Z. 2017, \aj, 153, 207, \dodoi{10.3847/1538-3881/aa683f}

\end{thebibliography}

\end{CJK*}
\end{document}